# Beyond Newtonian Mechanics: An Exploration of Active Force in Complex Swing Motions

Rong Li[1,2], Weicheng Cui[1,2*]

[1] Research Center for Industries of the Future, Westlake University, Hangzhou, Zhejiang 310030, China

[2] Key Laboratory of Coastal Environment and Resources of Zhejiang Province, School of Engineering, Westlake University, Hangzhou, Zhejiang 310030, China.

* cuiweicheng@westlake.edu.cn

## Abstract

Life mechanics, an emerging field, focuses on the self-organizing forces and motions within living systems. This study introduces the novel concept of 'active force,' generated by mind-body interactions, as an essential element in understanding self-organizing movements. We propose a new set of control equations to model the self-pumping swing motion by incorporating the active force into the Newton's second law. With this new mechanical framework, we analyzed the child's complex swing motions with rapid standing and squatting movements. The results reveal a pulse-like pattern of the active force along the swing length, driving changes in the radial speed and swing length. This active force counteracts the resistance and propels the swing, which is not attainable by the stone. Thus, active force underlies self-organization in living systems, warranting further exploration of its mechanics in life movements.

## 1. Introduction

Dyson, a physicist, remarked that the twenty-first century may be the century of biology [1]. Life, the most intricate of complex systems, is usually defined as a system that exhibits many nontrivial movements, including responsiveness, energy transformation, metabolism, growth, reproduction, and evolution [2]. Understanding these living movements poses the greatest challenge in modern science [3,4]. Accepting the axiom that force is the only reason for the change or movement of a body, Schrödinger's seminal question "What is life?" [5] can be reframed in the context of Newtonian mechanics: can we construct a mechanical model that describes nontrivial movements inherent in life?

Traditional Newtonian mechanics, which views the human body as the mechanical sum of its parts, overlooks the complexity of their interactions and emergent behaviors. From the late 19th century, it became clear that viewing life merely as a machine was insufficient for understanding phenomena, such as embryonic cell development. For instance, Driesch's experiments suggest that cells have an inherent ability to adapt to changing environments [6,7]. Further studies led to the birth of the modern system theory in the 1930s [8]. The "vitality" and "entelechy" postulated by early vitalists [9] found modern interpretation in the concept of "self-organization" in complex system science [10,11]. However, despite the development of numerous phenomenological differential equations and theories for complexities and lifes [11-15], mechanical descriptions of the dynamics of self-determined (i.e., by mind) movement in life remain rare. This void in physicists' understanding signals the need for the development of life mechanics [16].

In response to this need, we propose the concept of "active force," an internal force arising from mind-body interactions, as an integral part of life mechanics. Our new general system theory (NGST) [17-20] presents unified mechanics that incorporates both this active force from internal mind-body interactions and the passive forces arising from external interactions with other objects. Previous work on NGST has shown that the energy dissipated by resistance in a changing environment must be compensated by the active force [21,22]. In this study, we examined the dynamics of a simple pendulum system to illustrate the necessity of an active force to explain the observed phenomena in swing motions. By comparing the motion patterns of a child and stone of equivalent weight, we demonstrate that the active force underlies the self-determined movement unique to living systems. Our findings underscore the importance of incorporating an active force into Newton's second law as a fundamental paradigm in life mechanics.

# 2. General framework of Newton's second law with active force

Consider the mechanics of a particle within a multiparticle system situated in an Earth-fixed, noninertial coordinate system (as depicted in Fig.1). This model is grounded in the perception that our planet is in motion, an understanding that dates back to the era of Galileo Galilei. In this context, the governing equation for each particle is derived from Newton's second law:

$$\frac{d^2}{dt^2} m_i \boldsymbol{r}_i = \boldsymbol{F}_i^P + \boldsymbol{F}_i^A + \boldsymbol{F}_i^D, \tag{1}$$

where $m_i$ and $\mathbf{r}_i$ are the mass and displacement vectors of the $i$th particle, respectively, $\mathbf{F}_i^P$, $\mathbf{F}_i^A$ and $\mathbf{F}_i^D$ are the passive-driven force, active force, and dissipative force, respectively. Note that the vectors are indicated by Roman letters.

Eq. (1) embodies Newton's axiom that force is the agent of the motion change. The dynamics of nonliving objects can be adequately described by the passive-driven forces $\mathbf{F}_i^P$ and dissipation forces $\mathbf{F}_i^D$. These passive-driven forces were generated from other objects, including particles and the earth, in the system we studied. The most common passive-driven forces in the macroscopic world are gravity ($\mathbf{F}_i^G = GmM/r^2$) and the static electromagnetic force, that is, $\mathbf{F}_i^M = q_i(\mathbf{E} + \mathbf{v}_i \times \mathbf{B})$, where $\mathbf{E}$ is the electronic field, $\mathbf{B}$ is the magnetic field strength, $q_i$ and $\mathbf{v}_i = \dot{\mathbf{r}}_i$ are the charge and velocity of the particle, respectively. These forces can be expressed as a derivative of the generalized potential:

$$\boldsymbol{F}_i^P = -\nabla U + \frac{d}{dt}\left(\frac{\partial U}{\partial \dot{r}_i}\right), \tag{2}$$

where the generalized potential is $U = V_G + q\phi - q\mathbf{A} \cdot \mathbf{v}$. Here, $V_G$ is the gravity potential and $\phi$ and $\mathbf{A}$ are the scalar and vector potentials of the electromagnetic field, respectively. This analysis employs an Earth-fixed coordinate system and implicitly assumes the validity of Newton's second law in a noninertial coordinate system. According to the NGST ontology [21,22], we must abandon the assumption of inertial coordinate systems because they do not exist for human observers. However, the origin and expression of $\mathbf{F}_i^D$ are typically complex. The classical linear friction ($\mathbf{F}_i^D = -k_i \mathbf{v}_i$) and its corresponding Rayleigh dissipation function $D = \sum_i k_i \mathbf{v}_i^2 /2$ [23] is only a particular case (i.e., n = 1) of the general formula, that is, $\mathbf{F}_i^D = -k_i \mathbf{v}_i^n$. Although complex, friction is always defined along the inverse direction of the velocity.

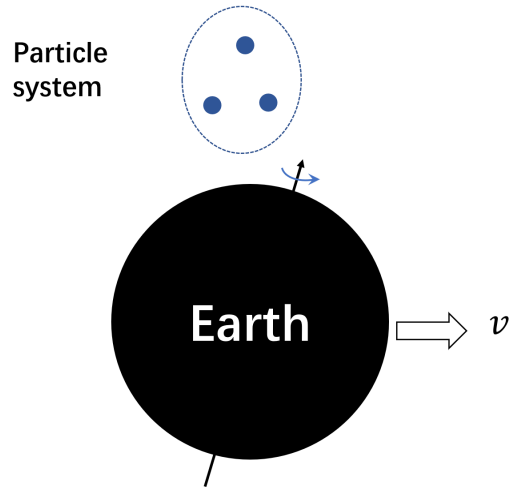

FIG. 1. A schematic representation of a N-body system in an earth-fixed coordinate system. It means that we only consider the motions of the particle systems relative to the observer on the earth.

In addition, a new type of force, referred to as the "active force," was introduced to explain the initiating movement changes in living entities. We adopted the distinction between active and passive forces using the force operator $F(B, C)$ introduced by Truesdell [24]. According to Truesdell, "a system of forces on a world $\Omega$ is an assignment of vectors in an inner-product space to all pairs of separate bodies of $\Omega$. The force vector $f(B, C)$ is called the force exerted on B by C." The four existing fundamental forces — gravitational, electromagnetic, strong, and weak — satisfy his definition when object B differs from object C. Hence, these four fundamental forces are designated as passive forces, corresponding to $f(B, C)$ with $B \neq C$. Besides, when object B is the same as object C, Truesdell defines the force $f(B, B)$, if it is non-zero, theories as a self-force (i.e., "the force exerted by B on itself"). However, Truesdell did not consider the physical properties of this self-force.

Thus, two immediate questions arise. Which objects possess self-force and how can this self-force be calculated? From our daily experiences, humans exhibit self-force during their self-determined movements. For example, humans and other animals can run by using force. To capture this phenomenon, we introduced an active force in our NGST ontology [17] that may answer the question of which objects can possess self-force. The answer is any entity containing a mind that can exert self-force; when the mind separates from the body, it signifies the death of the living body. After death, living objects no longer exhibited an active force. Therefore, active force is interpreted as a mind-body interaction. However, calculating this self-force is a significant challenge, as it is governed by unpredictable free will of the mind. Our suggestion is to first accept a model capable of explaining this phenomenon, and then gradually find a way to quantify this active force. This study presented an early effort and some initial results.

In general, everyday human movements such as walking, running, and swinging are inherently complex, necessitating rigorous modeling of the interactions and movements of various body parts [25-27]. However, this paper aims not to delve into the modeling of too much complex motions, but rather to clarify the differences between active forces in living systems and those in inanimate bodies. Therefore, our primary focus in this study was on the motion of the center of mass. In line with the definition of the center of mass coordinates ($\boldsymbol{R} = \sum_i m_i \boldsymbol{r}_i / \sum_i m_i$), Newton's second law can be obtained from the summation of Eq. (1) as,

$$M \frac{d^2}{dt^2} \boldsymbol{R} = \boldsymbol{F}_p + \boldsymbol{F}_a + \boldsymbol{f}, \qquad (3)$$

where $M = \sum_i m_i$ is the total mass, $\boldsymbol{F}_p = \sum_i \mathbf{F}_i^P$, $\boldsymbol{F}_a = \sum_i \mathbf{F}_i^A$, $\boldsymbol{f} = \sum_i \mathbf{F}_i^D$ are the total passive-driven force, total active-driven force, and the total dissipation force, respectively.

## 3. The active force for pumping a swing by a living system

This section explores how an active force propels a swing, particularly by comparing the dynamic differences (displacement, velocity, and force) between the swing motions of a child and stone. This comparison underscores the essential role of active force in explaining the motions of living systems.

## 3.1. The governing equation for pumping a swing

Applying Eq. (3) elucidates the differences between the swing motions of a child and stone, considering only simple pendulum motions within a vertical two-dimensional plane. Fig. 2 illustrates the simple pendulum system: (a) represents a classical case with a non-living stone, whereas (b) substitutes the stone with a child of equivalent weight. In the stone case, *L* denotes the length of the rigid massless rod, *m* denotes the mass of the stone, and *θ* represents the angle of the rod along the vertical axis. Furthermore, three forces acted at the center of the mass of the stone. First, gravity, $F_g = mg$ acts in the downward vertical direction. Second, a passive tension force, $\boldsymbol{T}_p$ acts along the rod owing to the balance of gravity and centrifugal force toward the frictionless pivot. Third, a friction force *f* resists the motion of the stone through the medium, which is assumed to be proportional to the velocity of the stone with a coefficient of friction *b* in the present work for simplicity. Hence, Newton's second law is expressed as follows:

$$m\frac{d\mathbf{v}}{dt} = \mathbf{F}_g + \mathbf{T}_p - b\mathbf{v}. \tag{4}$$

Because the passive tension force $\mathbf{T}_p$ is balanced by gravity and centrifugal force, the pendulum length is constant, that is, $L = L_0$. Therefore, it is straightforward to derive the equation for circumferential motion $v_\perp = L\dot\theta$ from Eq. (4) as

$$mL\ddot\theta(t) = -mg\sin\theta(t) - bL\dot\theta(t). \tag{5}$$

If we assume that the initial angle is very small, a general solution can be obtained as

$$\theta(t) = e^{-\frac{bt}{2m}}(c_+ e^{i\omega t} + c_- e^{-i\omega t}), \tag{6}$$

where $\omega = \sqrt{g/L_0 - b^2/4m^2}$ is the frequency of the damped pendulum, and $c_+$ and $c_-$ are two coefficients determined by the initial conditions. Here, the time-dependent coefficient $e^{-bt/2m}$ reveals that the pendulum motion is a damped oscillation that decays with time and finally ceases. That is, it is a passive motion that depends on the external pumping force.

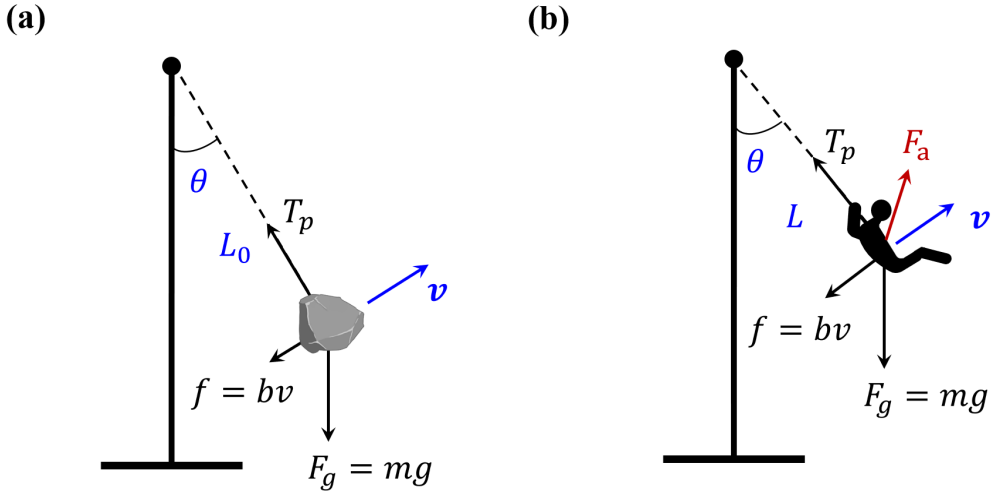

FIG. 2. (a) A force analysis of a simple pendulum of a lifeless stone. (b) A force analysis of a living pendulum, i.e., a child named Bob.

Conversely, Fig. 2 (b) depicts a child, Bob, as the pendulum, where the system behavior is influenced not only by external driving and damping forces, but also by an active force $\mathbf{F}_a$ acting in a distinct direction. This introduces a fundamental distinction between living beings and inanimate stone. In a prior analysis of an inanimate pendulum, the system was considered to be closed. However, Bob, being a person, introduces potential matter and information exchange with the environment through breathing, drinking, eating, and excreting. Nonetheless, focusing solely on Bob's swing motion allows us to overlook his matter and information exchanges with the environment. This simplification reveals four forces depicted in Fig. 2 (b): the internal active force $\boldsymbol{F}_a$, external gravitational force $F_G = mg$, external tension force $T$ along the rod, and external friction force $f$.

Establishing a relationship between the active (internal) and passive (external) forces is crucial. The total force exerted on an object, $\boldsymbol{F}_{\text{total}}$, is defined as the sum of the active and passive forces ($\boldsymbol{F}_a$ and $\boldsymbol{F}_p$ respectively). Thus, $\boldsymbol{F}_a$ can be expressed as $\boldsymbol{F}_a = \boldsymbol{F}_{\text{total}} - \boldsymbol{F}_p$. For an inanimate pendulum, $\boldsymbol{F}_a = 0$ and $\boldsymbol{F}_{\text{total}} = \boldsymbol{F}_g + \boldsymbol{T}_p - b\boldsymbol{v}$, where $T_p$—representing the swing rope tension owing to the passive balance of gravity and centrifugal force, which is defined as $T_p = mg\cos\theta(t) + mv_\perp^2/L$.

For a living pendulum, $\boldsymbol{F}_a \neq 0$. The active force, symbolizing a mind-body interaction, bifurcates into two categories: one instigating mutual motion within subsystems of the living entity and the other driving the overall center of mass movement. The former originates from the interaction forces among subsystems, independent of external reaction forces, whereas the latter necessitates external reaction forces from the environment. In our scenario, where Bob is considered a living pendulum, the significance of the latter interaction is apparent, manifesting as the force between Bob and the swing. This implies that the active force within the human body incites an active reaction force, $\boldsymbol{F}_a$, catalyzing the active movement of the person's center of mass. Consequently, Newton's second law applied to Bob is:

$$m\frac{d\boldsymbol{v}}{dt} = \boldsymbol{F}_g + \boldsymbol{T}_p + \boldsymbol{F}_a - b\boldsymbol{v}. \tag{7}$$

Eq. (7) reveals that the tension force comprises both active and passive parts, which are induced by the internal force of Bob and the external force of gravity, respectively. The active force in Eq. (7) represents the swing's reaction to a person, which can be oriented in any direction, yet is constrained by a maximum limit owing to the finiteness of Bob's mind-body interaction strength. We examined a straightforward scenario in which a person can move vertically while maintaining a stable stance on the board. In this case, the relationship $T_p = mg\cos\theta(t)$ consistently applies. Conversely, in the extreme case where a person disengages from the board, $T_p$ is equal to zero. Furthermore, any supplementary force between the individual and swing originates from the active force. Consequently, Bob possesses the capacity to move radially (where $L$ is not a constant) and in circumferential directions (where $\theta$ is not a constant):

$$m\frac{dv_r}{dt} = m\ddot{L}(t) = F_a^{\parallel} - b\dot{L}(t), \tag{8}$$

$$m\frac{dv_\perp}{dt} = m\frac{d[L(t)\dot{\theta}(t)]}{dt} = -mg\sin\theta(t) - bL(t)\dot{\theta}(t) + F_a^{\perp}, \tag{9}$$

where $F_a^{\parallel}$ and $F_a^{\perp}$ are the active forces in the radial and circumferential directions, respectively.

In Eq. (8), the radial motion is solely driven by the active force. By stating $L(t) = L_0 + l(t)$ where $l(t) = 0$ at $t \leq 0$, we can simplify these two equations as follows:

$$\ddot{l}(t) = \frac{F_a^{\parallel}}{m} - b\dot{l}(t). \tag{10}$$

$$\ddot{\theta}(t) = -\frac{g\sin\theta(t)}{L_0 + l(t)} - \left[\frac{b}{m} + \frac{\dot{l}(t)}{L_0 + l(t)}\right]\dot{\theta}(t) + \frac{F_a^{\perp}/m}{L_0 + l(t)}. \tag{11}$$

Eqns. (10) and (11) govern the general behavior of a living pendulum, requiring a solution to a system of nonlinear, nonhomogeneous ordinary differential equations. The active force yields a non-trivial active motion compared with the motion of a non-living pendulum. Significant differences include the following: (1) Active motion can result in radial movement, while passive pendulum motion primarily permits circumferential movement. (2) Active motion can sustain or augment the circumferential motion amplitude, known as the body's self-pumping swing [26,28-33], whereas passive pendulum motion can only decrease and rest. Certainly, these active motion patterns rely entirely on the active force patterns, that is, the magnitudes and temporal patterns of $F_a^{\parallel}$ and $F_a^{\perp}$. However, calculating this active force from a specific mind-body interaction mechanism is challenging problem to be solved. At present, only the inversion method can ascertain the active force from the measured trajectory, assuming Newton's second law as a universal law for any dynamics, and the resistance can be determined from prearranged experiments. As rudimentary examples to investigate the relationships between swing motions and active force, we will examine two simple modes of human pumping swing devoid of circumferential active force ($F_a^{\perp} = 0$) in the subsequent two subsections. The underlying logic is to select an appropriate motion pattern specifying the person's vertical movement, that is, $l(t)$ is a function of $\theta(t)$, enabling us to solve $\theta(t)$ from Eq. (11). We then derived the radial active force $F_a^{\parallel}$ from Eq. (10).

Pumping a playground swing is a very popular sport, especially for children. Its dynamic mechanism has been the subject of research for more than half a century of research [26,28-33]. It can be considered as a coupled oscillator system composed of a swing and a human. Typically, there are two pumping strategies: pumping from a standing position (as depicted in Fig. 3) and a seated position. In the former instance, the person stands at the lowest point and crouches at the highest point during swing motion. Each standcrouch cycle enhances the swing amplitude. The analysis demonstrated that each crouch-stand cycle provides a swing with an energy boost from the rider. In the latter scenario, the person abruptly rotates their body around the end of the swing chain. The amplitude of the swing increases as these rotations elevate the rider slightly above the highest level.

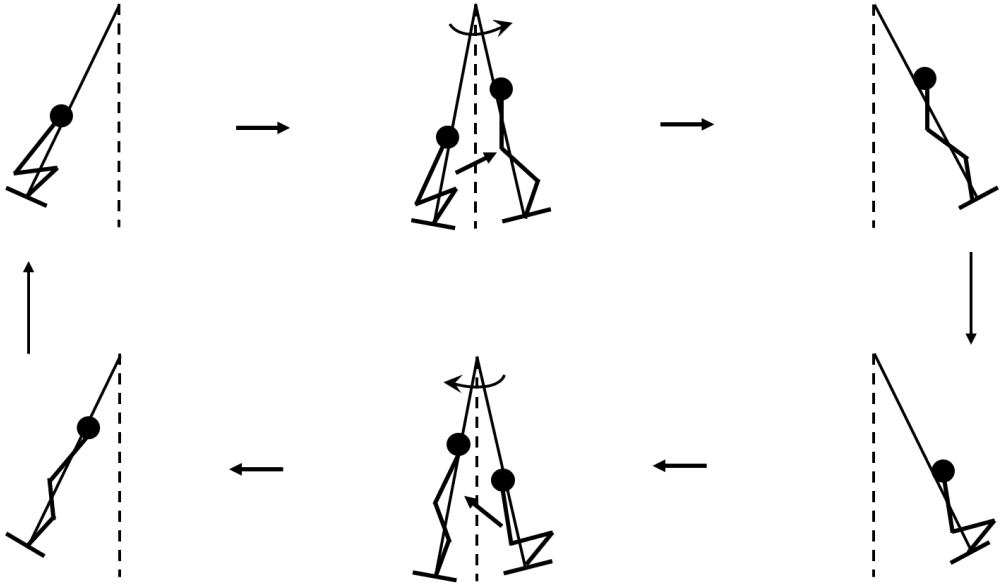

FIG. 3. Strategy for pumping a swing while standing, adapted from Ref. [34]. The child stands up near the lowest point and crouches down near the highest point during the swing motion.

It is worth noting that prior research on swing mechanics has primarily concentrated on effective pumping strategies, such as the pumping mode (standing or seated [30,34]) and the modulation of frequency and initial phase [33]. However, the driving force behind swing pumping has been less explored. This indicates that most studies have calculated the time evolutions of the angle and length, but not the pumping force. This force is generated by the human body and is referred to as the active force. If we focus on whether and how individuals can execute these movements, it is crucial to consider the temporal evolution pattern of this active force. Given that the force pattern corresponding to the stand-crouch motion is simpler, we mainly concentrate on pumping from the standing position in this study.

When pumping from a standing position, it can be presumed that only radial active motion is involved. Hence, the circumferential active force is zero, which is denoted by $F_a^\perp = 0$.

Consequently, the governing equations for a pump from the standing position are as follows:

$$F_a^{\parallel} = m[\ddot{l}(t) + b\dot{l}(t)], \tag{12}$$

$$\ddot{\theta}(t) = -\frac{g \sin \theta(t)}{L_0 + l(t)} - [\frac{b}{m} + \frac{\dot{l}(t)}{L_0 + l(t)}]\dot{\theta}(t). \tag{13}$$

To solve these equations, we must quantify the specific strategy of the standcrouch motions. Ideally, standing and crouching motions occur within a short time span. Therefore, a step function can be utilized to analytically represent abrupt changes in the pendulum length:

$$l(t) = -k[1 + 2H(\theta)H(\dot{\theta}) - H(\theta) - H(\dot{\theta})], \tag{14}$$

where $H(x) = 1$ $(x > 1)$ or $0$ $(x \leq 0)$ is a step (or Heaviside) function. Here, $H(\theta)$ characterizes the temporal pattern of standing up at the minimum angle, whereas $H(\dot{\theta})$ represents the temporal pattern of crouching down at the maximum angle. When $\theta > 0$ and $\dot{\theta} > 0$ or $\theta < 0$ and $\dot{\theta} < 0$, $l(t) = -k$, implying that the child is standing. When $\theta > 0$ and $\dot{\theta} < 0$ or $\theta < 0$ and $\dot{\theta} > 0$, $l(t) = k$, signifying that the child is crouching. The derivative of $l(t)$ can be calculated as:

$$\dot{l}(t) = -k[2\delta(\theta)H(\dot{\theta})\dot{\theta} + 2H(\theta)\delta(\dot{\theta})\ddot{\theta} - \delta(\theta)\dot{\theta} - \delta(\dot{\theta})\ddot{\theta}], \tag{15}$$

where $\delta(x)$ is the derivative of $H(x)$, equivalent to the Dirac function.

Eqns. (13)–(15) can only be solved using numerical simulations. To facilitate this, the Heaviside and Dirac functions must be substituted with finite-function approximations. Moreover, the actual stand-crouch motions transpire in finite time, necessitating a continuous time evolution function. To the best of our knowledge, one of the simplest approximations for the step function is the logistic (and pulse) function. Thus, for $H(\theta)$, $H(\dot{\theta})$ and their derivatives, we use

$$H(\theta) = \frac{1}{e^{-\frac{\theta}{c_a}} + 1}, \tag{16-a}$$

$$\delta(\theta) = \frac{d}{d\theta}H(\theta) = -\frac{1}{c_a}\frac{1}{2 + e^{-\frac{\theta}{c_a}} + e^{\frac{\theta}{c_a}}}, \tag{16-b}$$

$$H(\dot{\theta}) = \frac{1}{e^{-\frac{\dot{\theta}}{c_b}} + 1}, \tag{16-c}$$

$$\delta(\dot{\theta}) = \frac{d}{d\dot{\theta}} H(\theta) = -\frac{1}{c_b} \frac{1}{2 + e^{-\frac{\dot{\theta}}{c_b}} + e^{\frac{\dot{\theta}}{c_b}}}, \qquad (16-d)$$

Here, $c_a$ determines the temporal pattern of standing up at the minimum angle, whereas $c_b$ determines the temporal pattern of crouching down at the maximum angle. These two parameters can be independently chosen according to the child's free will.

## 3.2. The active force for pumping a swing without friction

By employing Eqns. (12)–(16), we can derive the time evolution of several parameters: the angle $\theta(t)$, active force $F_a^{\parallel}$, pendulum length $L(t) = L_0 + l(t)$, vertical height of the mass center of the child $h(t) = L_0 - L(t)\cos\theta$, the circumferential velocity $v_{\perp} = L(t)\dot{\theta}(t)$ and radial velocity $v_{\parallel} = \dot{l}(t)$. The angle and velocities in two directions depict the dynamics of the motion, whereas the alterations in swing length and height represent the spatial state. The active force, which is our primary objective, emerged from these calculations.

To execute this simulation, we must establish the parameters within these equations and initial conditions. Let us assume that the masses of the individual and stone are identical, that is, $m = 20$ kg. The swing's maximum pendulum length, $L_0$, was 2 m, and the gravitational acceleration $g$ was 9.81 m/s². We assume the height change of the child is $k = 0.3$m and zero for the stone. The friction coefficient $b$, is influenced by the friction between the individual and air, the swing, and between the swing rope and the fixed point. The coefficients $c_a$ and $c_b$, which dictate the patterns of standing and crouching, as well as the initial angle $\theta(0)$ and initial velocity $\dot{\theta}(0)$ can be freely chosen by the child's free will. Consequently, in this study, we treat parameters $b$, $c_a$, $c_b$, $\theta(0)$ and $\dot{\theta}(0)$) as free adjustable parameters.

By substituting these parameters into Eqs. (12)–(16), we can juxtapose the swings of the stone and the child. The simulations for the frictionless scenario are shown in Figure 4. As is commonly understood, the stone (represented by dashed black lines) performs harmonic oscillation, maintaining a constant amplitude throughout each period. Conversely, the child (denoted by solid red lines) generates an enhanced oscillation, with the amplitude increasing in each period. The figures show that the augmentation in the amplitudes of the angle (a) and circumferential velocity (b) aligns with the continuous increase in radial velocity (e), prompted by the child's periodically increasing active force (d). This active force emerged as four pulses per cycle, corresponding to two instances of standing and two instances of crouching. More specifically, during the first period, the initial angles (a) and initial circumferential velocities (b) for the stone and child were identical.

However, the child's swift standing motion around $t$=0.5-0.8 s (see (d)-(f)) leads to a gradual increase in the angle and the circumferential velocity, reaching peak differences from the stone around 1.5 s and 2.1 s. Consequently, we conclude that in swing motion, the child's active force indeed stimulates the amplification of the swing's amplitude.

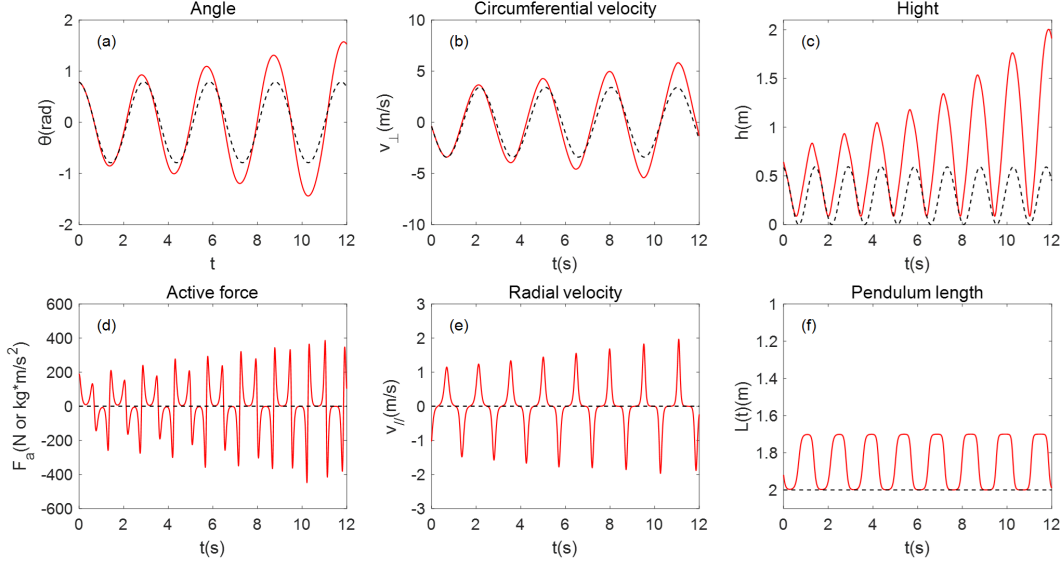

FIG. 4. Active-motion simulations for a child pumping a swing without friction, as depicted by the solid red lines. In contrast, the dashed black lines represent the simple pendulum motion of a stone with the same mass as the child. The parameters selected for these simulations are outlined in Table I.

Table I. Parameters selection for the stone (zero active force) and the child (finite active force) with zero friction.

| System | Mass | Maximum length | Gravitational acceleration | Friction coefficient | Standing height | Pulse angle | Pulse velocity | Initial angle | Initial velocity |
|---|---|---|---|---|---|---|---|---|---|
| | $M$(kg) | $L_0$(m) | g(kg. s$^{-2}$) | b(kg. s$^{-1}$) | $k$ | $c_a$ | $c_b(s^{-1})$ | $\theta(0)$ | $\dot{\theta}(0)$ |
| Stone | 20 | 2 | 9.8 | 0 | 0 | — | — | $\pi/4$ | $-0.2$ |
| Child | 20 | 2 | 9.8 | 0 | 0.3 | 0.12 | 0.20 | $\pi/4$ | $-0.2$ |

We proceeded with a quantitative analysis of the energy-pumping mechanism facilitated by the active force. In a frictionless environment, we assume that standing up and crouching down are executed instantaneously. Because the active force solely impacts the radial motion, neither the circumferential momentum (or velocity) nor the angle changes this process. The critical variables that vary are the pendulum length, which shifts from $L_0$ to $L_0 - k$, and the radial velocity $v_\parallel = \dot{l}(t)$. By examining the nth instance of standing up, we can establish a relationship between the maximum swing amplitudes before ($\theta_{n-1}$) and after ($\theta_n$) the actions. The law of energy conservation dictates that the circumferential kinetic energy ($m(v_\perp^n)^2/2$) at the lowest point originates from the potential energy decrease ($mgL_0(1 - \cos\theta_{n-1})$) of the child from the maximum swing angle $\theta_{n-1}$. After standing up, this kinetic energy transitions into a potential energy increase $(mg(L_0 - k)(1 - \cos\theta_n))$ of the maximum swing $\theta_n$. As such, the potential energy decrease before standing up is equal to the increase following it.

$$mgL_0(1 - cos\theta_{n-1}) = mg(L_0 - k)(1 - cos\theta_n). \tag{17}$$

From this relation, we can obtain the $n$th maximum angle, height, and velocity as:

$$\theta_n = \arccos\left[\frac{\cos\theta_{n-1} - k'}{1 - k'}\right], \tag{18 - a}$$

$$h_n = L_0 \frac{1 - \cos\theta_1}{(1 - k')^{n-1}}, \tag{18 - b}$$

$$v_\perp^n = \sqrt{2gL_0 \frac{1 - \cos\theta_1}{(1 - k')^{n-1}}}, \tag{18 - c}$$

where $k' = k/L_0$. Eq. (18) reveals that the swing amplitude increases every half-cycle according to a function related to $1/(1 - k')$. By substituting the parameters used in the numerical simulations into Eq. (18), we discover that $\theta_n = \arccos[1.18(\cos\theta_{n-1} - 0.15)]$, $h_n = 0.586 \times 1.18^{n-1}$, and $v_\perp^n = 3.39 \times 1.18^{(n-1)/2}$. These analytical expressions closely align with the numerical simulations. This consistency validates that the swing accumulates the net energy from the child's active motions in each pumping cycle. This net energy stems from the work done by the child's active force when standing up, which increases the height by k, over the energy spent when squatting down, which decreases the height by $k\cos\theta_n$.

## 3.3. The active force for pumping a swing with linear friction

Next, we explore a more realistic scenario in which both active and frictional forces are at play. A particularly noteworthy situation occurs when the active force counterbalances the frictional force, leading to stable oscillation with a constant amplitude that neither decays nor expands. Such a state represents a stable equilibrium that is frequently observed in swing sports. An intriguing question is whether our active-force model (Eq. (12) and (13)) can accurately simulate this condition. As shown in Fig. 5, the active-force model successfully replicates this scenario. Specifically, for the stone without the active force, the swing amplitude demonstrated a decay (as shown by the solid blue lines) when compared to the harmonic oscillations without friction (represented by dashed black lines). In contrast, for a child exerting an appropriately strong active force (denoted by solid red lines), the swing amplitude remains constant (as seen in images (a) to (c)). This essentially means that, in this situation, the energy pumped from the active force successfully counteracted the damping effect of friction.

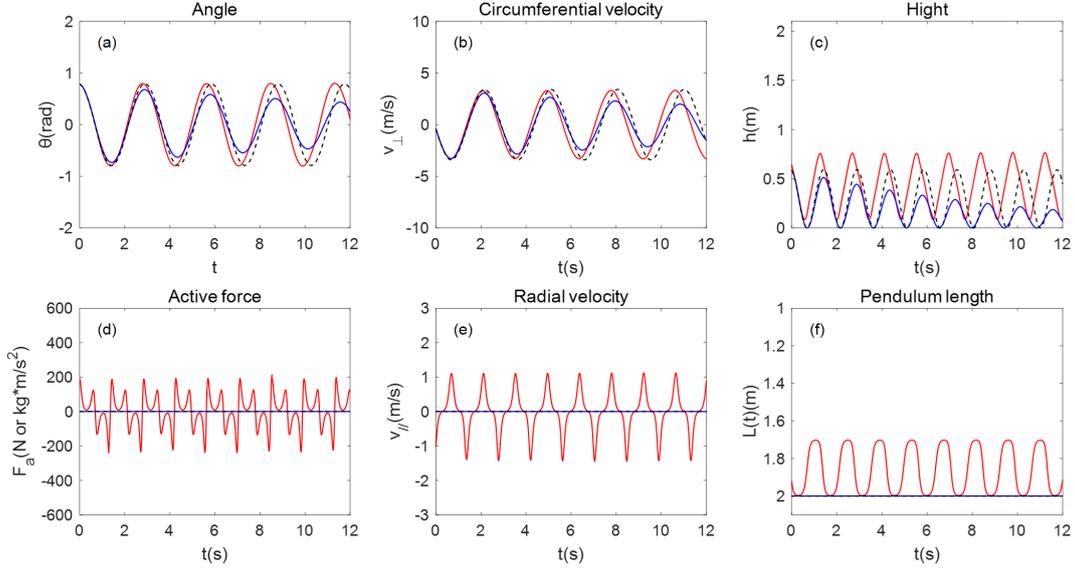

FIG. 5. Active-motion simulations showcasing a child pumping a swing in the presence of finite friction, as shown by the solid red lines. To serve as comparisons, the solid blue lines (indicating finite friction) and dashed black lines (representing no friction) illustrate the simple pendulum motion of a stone with the same mass as the child. The parameters used for these simulations are listed in Table II.

Table II. The parameters selected for the swing simulations involving a stone (zero active force) and a child (finite active force) under conditions of finite friction, as shown in Fig. 5.

| System | Mass | Maximum length | Gravitational acceleration | Friction coefficient | Standing height | Pulse angle | Pulse velocity | Initial angle | Initial velocity |
|---|---|---|---|---|---|---|---|---|---|
| | $M(kg)$ | $L_0(m)$ | $g(kg \cdot s^{-2})$ | $b(kg \cdot s^{-1})$ | $k$ | $c_a$ | $c_b(s^{-1})$ | $\theta(0)$ | $\dot{\theta}(0)$ |
| Stone | 20 | 2 | 9.8 | 2 | 0 | — | — | $\pi/4$ | $-0.2$ |
| Child | 20 | 2 | 9.8 | 2 | 0.3 | 0.12 | 0.20 | $\pi/4$ | $-0.2$ |

Initially, it can be inferred that the frictional force is counterbalanced by the active force, implying that only the net active force minus the frictional force needs to be considered. However, this was not the case. As is evident from Fig. 5, notable differences exist between the balance of the active force and frictional contributions to the swing amplitude (solid red lines) and the simple harmonic motion, where both forces perfectly nullify each other (dashed black lines). These discrepancies are manifested in the smaller period of the former case (2.84 s) by 4% compared with the latter case (2.95 s), despite a slightly larger amplitude in the former case (0.805±0.011 rad) by 2% compared with the latter case (0.790±0.002 s). This is intriguing, as a larger amplitude generally corresponds to a larger period in a scenario devoid of an active force. Nevertheless, according to the simple relation $T = 2\pi\sqrt{L/g}$, the nontrivial smaller period for the child is brought about by the reduction in the pendulum length owing to the standing squatting motions. The root cause of this is the frictional force comprising both the radial and circumferential components, in contrast to the active force, which is exclusively radial. This results in the impracticality of perfectly counteracting the frictional force regardless of the form of the radial active force. The actual scenario involves achieving an energy balance with the radial work executed by the active force compensating for the energy loss induced by the circumferential component of the frictional force.

In this context, a significant question is the quantitative conditions that constitute the energy balance between the active and frictional forces during a steady swing. Intuitively, the increase in amplitude (maximum angle or height) or energy driven by the active force over a cycle should be equal to the decrease in amplitude or energy caused by resistance. Taking height as an example, we propose that the increase in height attributed to the active force should be commensurated with the decrease in height caused by the frictional force, that is, $\delta h_+ \approx \delta h_-$. To analytically understand this equilibrium condition, we can approximate these two types of height changes by considering swings influenced solely by the active force (absence of frictional force, referred to as red lines in Fig. 4) and those influenced solely by the frictional force (absence of active force, referred to as blue lines in Fig. 5). From Eq. (18-b), we can ascertain that the height increases during each standing action with the active force and without friction as

$$\delta h_+ = L_0(\cos\theta_{n-1} - \cos\theta_n) = L_0 \frac{k'}{1-k'}(1 - \cos\theta_{n-1}). \tag{19}$$

Simultaneously, considering the influence of resistance in the absence of an active force, the analytical solution at this time is $\theta(t) = e^{-bt/2m}(ae^{i\omega t} + be^{-i\omega t})$. Ignoring the influence of friction on frequency (or period $T = 2\pi/\omega$), then $\omega = \sqrt{g/L_0}$. Then, the decay coefficient of the maximum angle caused by damping within a half period is $\exp[\pi b/(2m\sqrt{g/L_0})]$. The corresponding decrease in height is:

$$\delta h_- = L_0(\cos\theta_n - \cos\theta_{n-1}) \approx L_0 \cos\theta_{n-1} \sin\theta_{n-1}\, \theta_{n-1}\left(1 - e^{-\frac{\pi b}{2m}\sqrt{\frac{L_0}{g}}}\right). \tag{20}$$

Substituting the parameters from Table II, we obtained $\delta h_+ = 0.103$ m and the $\delta h_- \approx 0.538$ m at $n = 2$. Although the two height changes are not identical, they are of the same order of magnitude, rendering $\delta h_+ \approx \delta h_-$ a useful approximate condition for the energy balance analysis of a real, steady swing.

We now turn to the primary parameters that determine the amplitude of the swing propelled by the active force. Holding constant parameters such as the child's mass, swing length, resistance, and initial conditions, we found that the child's center of mass elevation (the height difference between the standing and squatting positions) significantly influences the amplitude. For instance, Fig. 6(a) illustrates that when the standing amplitude surpasses 0.3 meters, the oscillation of the swing gradually increases (as indicated by the blue lines). Conversely, when the amplitude drops below 0.3 meters, the oscillation diminishes progressively (indicated by dashed black lines). In contrast, when the difference in the center of mass height between standing and squatting is fixed at 0.3 meters for each swing (Fig. 6f), the speed of standing ($c_a$) and squatting ($c_b$) barely affects the amplitude (Fig. 6d), despite the substantial variation in the peak active force. Therefore, beyond the typical factors influencing both animate and inanimate systems, such as mass, swing length, resistance, and initial conditions, the pivotal determinant of the amplitude in a swing propelled by standing is the difference in the center of mass height between standing and squatting.

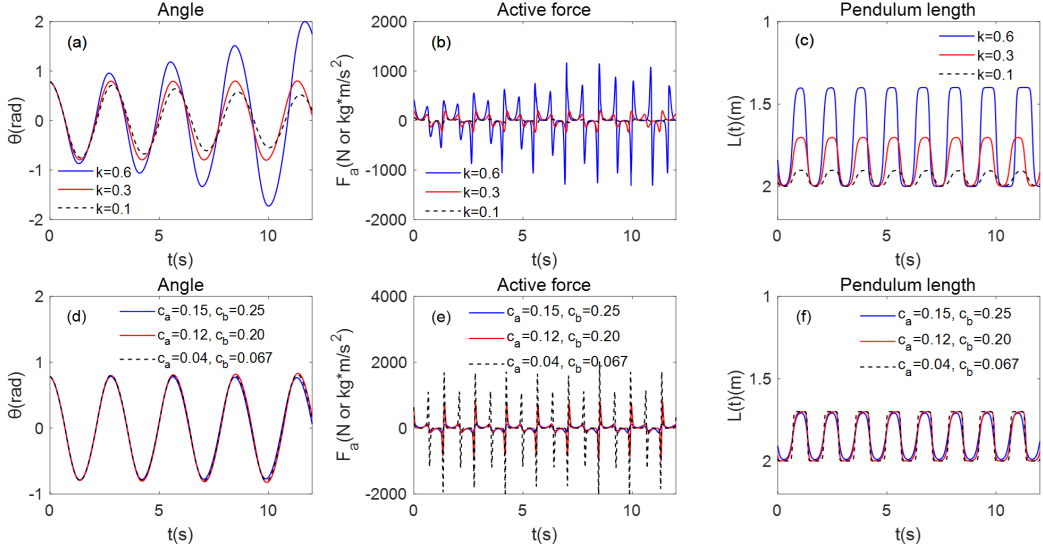

FIG. 6. The amplitude alterations attributable to height (k) variation ((a)-(c)) or speed ($c_a$ and $c_b$) changes in the standing-squatting motion ((d)-(f)). The adjusted parameters are denoted in legends. Other parameters align with those specified for active motions in Table II.

## 4. Discussion and conclusions

This paper introduces a compelling argument for the existence and necessity of the 'active force'

concept to elucidate the dynamics of living systems, using the human-pumping swing action as an example. This active force, originating from the interaction between the body and mind, provides a unique contribution to the dynamics of living systems, a facet that conventional physics often disregards. By incorporating an active force into Newton's second law, we established a comprehensive mechanical framework applicable to both living and nonliving systems. This innovative governing equation was applied to describe the swing motions of both a child and an equally weighted stone, with a particular focus on their differences. Our study reveals that children possessing active force can sustain or even augment their amplitude despite friction, whereas a stone's oscillation can only decrease. Quantitative analysis indicated that the energy pumping caused by the active force performing work in each standing motion was responsible for this phenomenon. Furthermore, we identified that the active force exhibits a pulse-like pattern, whereas a linear style (e.g., $l = -k\theta$) fails to generate a self-excited oscillation (as detailed in the Appendix). These findings underscore the existence and necessity of the 'active force' concept for understanding living system dynamics, offering a paradigm and mechanical framework for further exploration into life mechanics [16].

However, this research constitutes a preliminary study of active force. Future research should first probe further into the quantitative measurement and calculation of the active force in the self-determined motions of living beings, such as development, metabolism, and sports in plants, animals, and humans. Such efforts may reveal novel patterns and mechanisms of active force, laying a scientific foundation for predicting and controlling life movements. It is crucial to remember that the active force discussed herein represents the summation of all the active forces experienced by the human body's center of mass. Actual measurements might necessitate measuring the contact at multiple points between the human body and the interacting object, such as the contact of the feet and hands with a swing. By vectorially synthesizing the forces at multiple points, we can obtain the resultant force at the center of mass, as discussed in this paper. Consequently, further theoretical development should consider the movement of multiple body parts under forces acting at various points on the human body via Eq. (1). In this context, an immediate question would be to investigate the active force pattern in 'pumping a swing from a seated position' and compare it with the results of the present study. This exploration would necessitate considering three subsystems of the human body: the torso, thighs, and lower legs [25,31].

Moreover, the relationships between active force, bioelectricity, and consciousness present an intriguing topic. The active forces shown in Fig. (4)–(6) exhibit a pulsating characteristic in the time series, reminiscent of the pulsation patterns of bioelectric signals such as ECG and EEG [35]. We hypothesized that this similarity is not coincidental, suggesting that active force indeed embodies the characteristics of life. This is because the active force pattern in the pump swing is a consequence of an individual's conscious decision in their movement mode, which is a product of the brain and heart electrical activities. In future studies, we will propose simultaneous measurements and mathematical modeling of active force with EEG and ECG signals to investigate their quantitative correlations, potentially advancing our understanding of the underlying physical mechanisms connecting active force, bioelectricity, and human conscious movement.

## Acknowledgment

This research was funded by the National Key Research and Development Program of China (2022YFC2805200), start-up funding from Westlake University (041030150118), and Scientific Research Funding Project of Westlake University (2021WUFP017). We thank OpenAI for the GPT-4 model, which helped us improve the language of the manuscript.

## Data Access Statement

All relevant data are within the paper.

## Conflict of Interest declaration

The authors have declared that no competing interests exist.

## Appendix The active force for a linear response to the swing

This section demonstrates that the linear pumping mode fails to generate self-excited oscillations. This is the most straightforward scenario involving a linear relationship between the radial and circumferential movements, that is, $l(t) = -k\theta(t)$. This suggests that the body manipulates limb movements to decrease the pendulum length as the center of mass ascends, and increases it as the center of mass decreases. Concurrently, we postulate that at this juncture, the circumferential active force applied by the individual is negligible relative to the radial active force, i.e., $F_a^\perp \ll F_a^\parallel$. Under these circumstances, the following equations can be derived.

$$\ddot{\theta}(t) = -\frac{g \sin \theta(t)}{L_0 - k\theta(t)} - [\frac{b}{m} + \frac{-k\dot{\theta}(t)}{L_0 - k\theta(t)}]\dot{\theta}(t). \qquad (A1)$$

$$F_a^\parallel = -km\ddot{\theta}(t) - kb\dot{\theta}(t) = k\frac{mg \sin \theta(t) - mk[\dot{\theta}(t)]^2}{L_0 - k\theta(t)}. \qquad (A2)$$

Eq. (A2) represents the "exact" active force under the condition $l(t) = -k\theta(t)$, without any further approximations, and is a function of gravity, angular velocity, and pendulum length. To facilitate physical understanding, we can examine the simplest case using an analytical solution. Considering the small angle scenario, i.e., $\theta(t) \ll 1$, $[\dot{\theta}(t)]$ is the second order of $\theta$, so the active force can be approximated as

$$F_a^{\parallel} \approx \frac{k}{L_0} mg \sin \theta(t) \approx \frac{k}{L_0} mg \theta(t), \tag{A3}$$

, which is proportional to the restoring force along the circumference. Concurrently, Eq. (A1) can be simplified as

$$\ddot{\theta}(t) \approx -g'\theta(t) - b'\dot{\theta}(t) - g'k'\theta^2(t) + k'[\dot{\theta}(t)]^2 + (k')^2 \theta(t)[\dot{\theta}(t)]^2. \tag{A4}$$

where $g' = g/L_0$, $k' = k/L_0$, and $b' = b/m$. If we hypothesize that the active force is exceedingly small, then $k'$ can be entirely disregarded, which returns to the straightforward scenario of damped oscillation, where $\theta(t) \approx e^{-\frac{bt}{2m}}(c_+ e^{i\omega t} + c_- e^{-i\omega t})$. Consequently, the general solution for the active force is:

$$F_a^{\parallel} \approx \frac{k}{L_0} mg e^{-\frac{bt}{2m}}(c_+ e^{i\omega t} + c_- e^{-i\omega t}). \tag{A5}$$

At this point, the active motion of the human body is only a slight radial swing, and its active force is a function of the damped oscillation, which is not sufficient to overcome the resistance. This is usually the case for people who are not good at swinging. For large-angle movements, we can conduct a numerical simulation based on Eq. (A1) and (A2), as shown in Figure 7. The results indicate that a simple linear strategy, $l(t) = -k\theta(t)$, cannot amplify the amplitude of the swing.

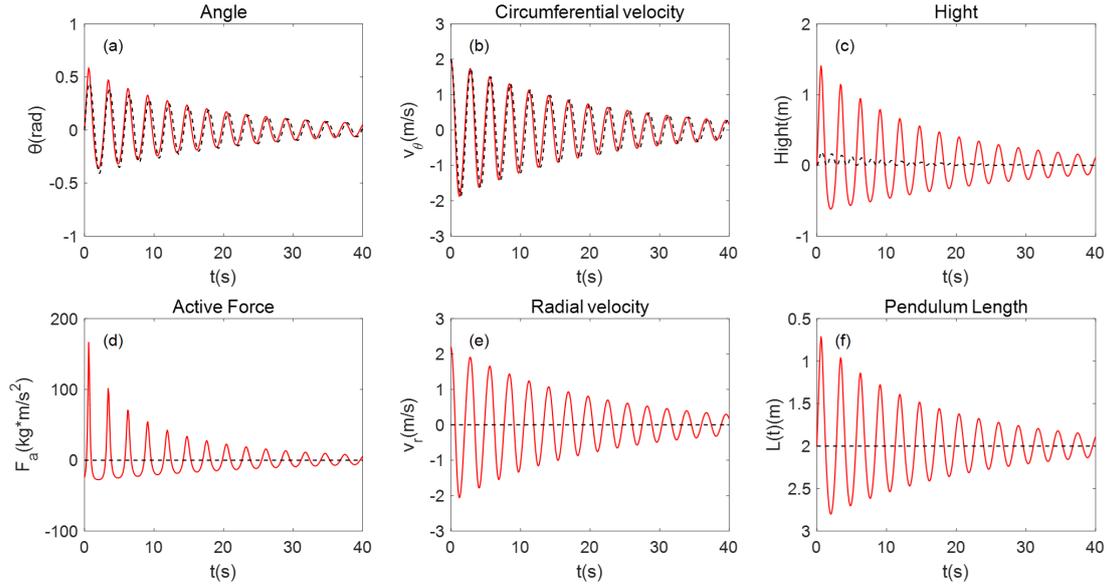

FIG. 7. Simulations for linear response type of active motion, as shown by the solid red lines. To serve as comparisons, the dashed black lines illustrate the simple pendulum motion of a stone with the same mass as the child. The parameters used for these simulations are listed in Table II.

Table III. Parameters' selection of the swing simulations for the stone (zero active force) and the child (finite active force) with finite friction are shown in Fig. 7.

| System | Mass | Maximum length | Gravitational acceleration | Friction coefficient | Standing height | Pulse angle | Pulse velocity | Initial angle | Initial velocity |
|---|---|---|---|---|---|---|---|---|---|
| | $M$(kg) | $L_0$(m) | g(kg·s$^{-2}$) | b(kg·s$^{-1}$) | $k$ | $c_a$ | $c_b(s^{-1})$ | $\theta(0)$ | $\dot{\theta}(0)$ |
| Stone | 20 | 2 | 9.8 | 2 | 0 | — | — | 0 | 1 |
| Child | 20 | 2 | 9.8 | 2 | 2.2 | — | — | 0 | 1 |